\documentclass{myaa}

\usepackage[varg]{txfonts}

\usepackage{natbib}
\usepackage{graphicx}
\usepackage{epstopdf}
\usepackage{amssymb}
\usepackage{amsmath,esint}
\usepackage{float}

\begin{document}

\title{On the formation of our solar system and many other protoplanetary systems observed by ALMA and SPHERE}

\author{Dimitris M. Christodoulou\inst{1,2}  
\and 
Demosthenes Kazanas\inst{3}
}

\institute{
Lowell Center for Space Science and Technology, University of Massachusetts Lowell, Lowell, MA, 01854, USA.\\
\and
Dept. of Mathematical Sciences, Univ. of Massachusetts Lowell, 
Lowell, MA, 01854, USA. \\ E-mail: dimitris\_christodoulou@uml.edu\\
\and
NASA/GSFC, Laboratory for High-Energy Astrophysics, Code 663, Greenbelt, MD 20771, USA. \\ E-mail: demos.kazanas@nasa.gov \\
}


\def\gsim{\mathrel{\raise.5ex\hbox{$>$}\mkern-14mu
                \lower0.6ex\hbox{$\sim$}}}

\def\lsim{\mathrel{\raise.3ex\hbox{$<$}\mkern-14mu
               \lower0.6ex\hbox{$\sim$}}}

\abstract{
In view of the many recent observations conducted by ALMA and SPHERE, it is becoming clear that protoplanetary disks form planets in narrow annular gaps at various distances from the central protostars before these protostars are actually fully formed and the gaseous disks have concluded their accretion/dispersal processes. This is in marked contrast to the many multi-planet exoplanetary systems that do not conform to this pristine picture. This major discrepancy calls for an explanation. We provide such an explanation in this work, based on analytical solutions of the cylindrical isothermal Lane-Emden equation with rotation which do not depend on boundary conditions. These ``intrinsic'' solutions of the differential equation attract the solutions of the Cauchy problem and force them to oscillate permanently. The oscillations create density maxima in which dust and planetesimals are trapped and they can form protoplanetary cores during the very early isothermal evolution of such protoplanetary nebulae. We apply this model to our solar nebula that formed in-situ a minimum of eleven protoplanetary cores that have grown to planets which have survived undisturbed to the present day. We are also in the process of applying the same model to the ALMA/DSHARP disks.}


\authorrunning{ }
\titlerunning{Formation of solar system/protoplanetary systems}

\maketitle

\section{Introduction}\label{intro}

\subsection{Planet Arrangements in Multi-Exoplanetary Systems}\label{intro1}

Discoveries of exoplanets started with \cite{may95} and \cite{mar96} and in the subsequent 24 years we ended up finding thousands of planets, many of them in multi-exoplanetary systems (\url{ www.exoplanets.org}). None of the systems containing several planets resembles our own solar system. With this statement, which has also been made by others in the field \citep[e.g.,][]{izi18,ray18}, we take exception to the many layman articles claiming that exoplanetary systems have been found that are similar to our own ({\it Science Daily}, 7/4/2003; {\it Universe Today}, 8/24/2010 and 11/27/2013; {\it Space.com}, 7/25/2012; {\it Cosmos}, 8/3/2015;  {\it The Verge}, 2/22/2017). The reason for such ``fake news'' is that professional astronomers have not voiced their opinions loudly, and the reason for their reluctance is probably that the ``truth'' is utterly disturbing: once again, we are facing a severely unpalatable conclusion, that our solar system is special, and we all know that this cannot be right.

\cite{ray18} recently  tried to compile and study some exoplanetary systems that, like our own, have a Jupiter-mass planet at a distance of about 5 AU. In the end, they concluded that our solar system is ``weird'' and it does not resemble any of the systems currently known. They and \cite{izi18} then argued that our solar system has been sculpted by multiple rare processes that could not successfully combine in any other known system. This premise is unacceptable because it reinforces the point made above about our solar system being special.  Furthermore, as we shall see below, it has been invalidated by the recent ALMA/DSHARP and SPHERE observations \citep[][to name a few]{alm15,mac18,ave18,cla18,kep18,guz18,ise18,zha18,kud18,lon18,pin18}.

The right approach to the problem is not to find ways to justify what we all know to be false; instead, we should be trying to understand (a) why protoplanets did not migrate in the solar nebula (unlike in most exosystems), and (b) why we have not yet observed some exosystems that resemble our own. We believe that we already have the answers to these two questions, answers that are not known to many or they have been overlooked by the few.

\subsection{The Recent ALMA and SPHERE Observations}

The answers to the above questions cannot probably come from planet searches because the sample of nearby stars accessible to observations is severely distance and volume limited. In addition, these systems are fully formed, the gaseous disks are gone, and there has been plenty of time for instabilities to cause migration and destruction of any pre-existing planets in well-organized arrangements, such as the planets in our solar system. On the other hand, ALMA and SPHERE observations are subject to the same limitations to a much lesser degree, they are observing much younger systems, and they can map out young protoplanetary disks with resolutions down to 1~AU. The beautiful images shown in \cite{alm15}, \cite{ave18}, and \cite{cla18} show relatively thick disks orbiting young or not-fully-formed protostars with planets having already formed and having carved out many annular gaps in the gas.

We point out that the radial brightness maps for CI Tau shown in \cite{cla18} and those shown by ALMA/DSHARP \citep{and16,rua17,dul18,fav18,guz18,hua18,ise18,per18,zha18} resemble the radial density profiles that we have also obtained for our theoretical model nebula, which we describe below.

\subsection{Isothermal Lane-Emden Equation with Rotation}

We have used a new technique to solve analytically the Lane-Emden equation \citep{lan69,emd07} for an isothermal, self-gravitating gas in cylindrical symmetry. The analytical solutions are \textit{singular} and \textit{intrinsic} to the equation itself, as they do not depend on or reproduced by any particular boundary conditions. In simple terms, these solutions are favored by the differential equation which has no regard for any boundary conditions that may be imposed externally by the Cauchy problem. When the Cauchy problem is solved numerically, its physical solutions are unable to match the preferred intrinsic solution (because of the imposed boundary conditions), but they are attracted to it, and they are forced to oscillate permanently about this fundamental solution.  Such mandatory oscillations develop both in the radial density profile and in the gravitational potential. The model incorporates in its input a large family of differential rotation profiles, and differential rotation dictates the spacing and radial extent of the multiple local potential wells. These are approximately equidistant only for the case of uniform rotation. The model lacks scale, but we can remedy this problem in solar-system modeling by fitting the locations of the planets to successive radial potential minima and by setting the distance of the third minimum from the center equal to 1~AU (planet Earth).

Applied to the very early isothermal solar nebula, this model predicts that planets ought to form safely inside consecutive potential minima from which dust and pebbles could not escape, unlike dissipative gas that continues to inflow straight toward the center to form the protosun. The central mass is a lot lower than 1$M_\odot$, so we predict that protoplanetary cores ought to form long before the protosun. This prediction, also made by \cite{gre10}, has recently received strong support from the observations of the still-forming star TMC1A \citep{har18} that is only about 0.1 Myr old.

One objection to this model has been that it is cylindrical. We do not believe this is a problem because the ALMA-SPHERE disks are not geometrically thin Keplerian disks \citep[see][]{lee17,lee18,zha18} since the protostars have not fully formed and the gaseous self-gravitating disks are relatively thick and massive. In addition, we also subscribe to an old notion put forth by Sir James \cite{jea14} last century: \textit{``All the essential physical features of the natural three-dimensional problem appear to be reproduced in the simpler cylindrical problem, so that it seems legitimate to hope that an
argument by analogy may not lead to entirely erroneous result.''}\,\footnote{\cite{jea14} was solving the cylindrical Lane-Emden equation with uniform rotation when he made these remarks. The solution turned out to be a zeroth-order Bessel function oscillating about the intrinsic solution $\tau = \beta_0^2$ (see below). It took us more than 100 years to do better than that and solve analytically differentially rotating Lane-Emden systems.}

Another objection to this oscillatory model stems from a misunderstanding of its application regime. The isothermal nebular model described below does not apply to the present state of the solar system or the ALMA disks that have undergone evolution for millions of years and the gas has been heated by various processes \cite[][and references therein]{toh02}. It only applies to the early isothermal evolution of their nebulae and long before any heating processes and ionization set in. For example, we derive in \S~\ref{rhomax} maximum typical densities of molecular hydrogen and neutral helium of $\sim 4\times 10^{-9} {\rm ~g} {\rm ~cm}^{-3}$ in the core of our model solar nebula, values that are roughly consistent with the extended isothermal phase of the solar nebula \citep{toh02}. Our view then is that extrasolar protoplanetary cores were formed during the same very early isothermal phase of evolution, and the solar-system planets and the observed dark gaps in present-day ALMA disks have survived ever since, as there were no resonant interactions or self-gravity induced instabilities in these disks capable of destroying their pristine, well-organized structures in the subsequent 0.1 Myr - 4.5 Gyr.

\subsection{Outline}

In \S~\ref{theory}, we describe the intrinsic solutions of the Lane-Emden equation and the resulting model of the solar nebula. In \S~\ref{apps}, we apply the model to the dark disk gaps observed by ALMA and believed to be the sites of already formed planets. In particular, we search for unstable dark gaps in mean-motion resonances and/or in a Titius-Bode arrangement. These empirical phenomena weigh heavily on to the long-term stability of such still-forming exoplanetary systems. Finally, in \S~\ref{disc}, we summarize our results. More detailed modeling of ALMA disks is also currently under way.

\section{Intrinsic and Oscillatory Solutions of the Isothermal Lane-Emden Equation with Rotation}\label{theory}

\subsection{Isothermal Equilibrium Models}\label{models}

We consider the axisymmetric equilibria that are available to a
rotating self-gravitating gas in the absence of viscosity and
magnetic fields. We adopt cylindrical coordinates ($R, \phi, z$) and
the assumption of cylindrical symmetry ($\partial /\partial z = 0$)
which lets us ignore $z$-dependent gradients and reduces the problem
to one dimension, the distance $R$ from the rotation axis. This
technique has become common practice in studies of rotating,
self-gravitating, fluid disks \citep[e.g.,][]{goo88,chr92,chr96} because it simplifies the stability
analyses of effectively two-dimensional modes of disturbance. In what
follows, we are interested in equilibrium structures
that describe the physical
conditions close to the midplane of a geometrically thick protoplanetary disk, so the
assumptions $\partial /\partial\phi = 0 = \partial /\partial z$ allow
us to tackle the problem by solving ordinary differential equations (ODEs).

We further adopt a rotation law of the form
\begin{equation}
\Omega (R) = \Omega_0\cdot f(x)\, ,
\label{rot}
\end{equation}
where $x\equiv R/R_0$ is a dimensionless radius and the scale length $R_0$ 
will be specified in eq.~(\ref{length}) below. Furthermore, $\Omega (R)$ is the angular 
velocity, $\Omega_0$ is the value of $\Omega$ at some fixed radius, and the
dimensionless function $f(x)$ for differential rotation is generally an
arbitrary function of $x$. For centrally condensed models, it is
convenient to choose $\Omega_0 = \Omega (0)$ and the regularity condition
$f(0)=1$.

Finally, we assume an isothermal equation of state of the form
\begin{equation}
P = c_0^2\cdot \rho\, ,
\label{state}
\end{equation}
where $P$ is the thermal pressure, $\rho$ is the gas density, 
and $c_0$ is the constant isothermal sound speed.

\subsection{The Lane-Emden Equation With Rotation}\label{LER}

Axisymmetric and cylindrically symmetric, nonmagnetic equilibria for a perfect fluid 
are described by the equation of hydrostatic equilibrium
\begin{equation}
\frac{1}{\rho}\frac{dP}{dR} + \frac{d\Phi}{dR} = \Omega^2 R\, ,
\label{hydro}
\end{equation}
where the gravitational potential $\Phi (R)$ satisfies Poisson's equation
\begin{equation}
\frac{1}{R}\frac{d}{dR}R\frac{d\Phi}{dR} = 4\pi G\rho\, ,
\label{poisson}
\end{equation}
where $G$ is the Newtonian gravitational constant. Combining
eqs.~(\ref{rot})-(\ref{poisson}) and using the definition $x\equiv R/R_0$, 
we find a second-order nonlinear innonhomogeneous ODE that can be cast in 
the form
\begin{equation}
\frac{1}{x} \frac{d}{dx} x \frac{d}{dx}\ln\tau \ + \ \tau \ = \ 
\frac{\beta_0^2}{2x}\frac{d}{dx}\left(x^2 f^2\right)\, ,
\label{main1}
\end{equation}
where ~$\tau\equiv\rho /\rho_0$, ~$\rho_0$ is the maximum 
density or a fixed cutoff density for singular/annular models,
~$\beta_0\equiv \Omega_0 /\Omega_J$, ~$\Omega_J^2\equiv 2\pi G\rho_0$, ~and
\begin{equation}
R_0^2 \equiv \frac{c_0^2}{4\pi G\rho_0} = \frac{c_0^2}{2\Omega_J^2}\, .
\label{length}
\end{equation}
The term $\Omega_J$ represents the gravitational (Jeans) frequency and 
the dimensionless rotation parameter $\beta_0$ measures centrifugal
support against self-gravity; in general, $0 \leq \beta_0 < 1$, since the 
gas is also partially supported by pressure gradients in the radial direction.

Eq.~(\ref{main1}) reduces to the classical isothermal Lane-Emden equation 
in the absence of rotation ($\beta_0 = 0$).\footnote{And for a flat rotation curve of the form $f(x)=1/x$.} 
In the following subsection, we derive analytically
a class of particular solutions of the general 
problem (eq.~(\ref{main1}) with an enormous family $f(x)$ of differential rotation profiles).

\subsection{Differentially Rotating Solutions}\label{withrot}

When the right-hand side (hereafter RHS) of eq.~(\ref{main1}) is nonzero
(i.e., when $\beta_0\neq 0$ and $f(x)\neq 1/x$), the property
of scale invariance is lost from all cases of interest (uniform rotation,
power-law rotation, etc.), irrespective of the prescription chosen for the 
differential rotation function $f(x)$.\footnote{Eq.~(\ref{main1}) with a nonzero
RHS is scale invariant only for $f(x) = \sqrt{A\ln x + B}/x$, where $A$ and $B$
are arbitrary constants. This case can be solved by transforming the scale-invariant
ODE to its autonomous form, but there is no need
to do so; the same solution is obtained easier by the method described in this
subsection.} 
Then, eq.~(\ref{main1}) has no special symmetry associated with it, 
and this is probably why 
some interesting features that we describe below have gone unnoticed for so long.

The RHS of eq.~(\ref{main1}) is not merely a rotation-dependent correction term to the 
classical isothermal Lane-Emden equation. The introduction of rotation 
changes the properties of the ODE to such a large extent that the known nonrotating solutions \citep{sto63,ost64} cannot guide the effort to find rotating equilibrium 
solutions. (This will be evident in Figs.~\ref{fig2} and~\ref{fig3} below.)
In fact, it is the functional form of the RHS that determines now the preferred (intrinsic) 
solutions of the ODE: By equating the last two terms of eq.~(\ref{main1}), we can write down 
an entire class of particular equilibrium solutions, viz.
\begin{equation}
\tau (x) \ = \ \frac{\beta_0^2}{2x}\frac{d}{dx}\left(x^2 f^2\right)\, ,
\label{part1}
\end{equation}
provided that
\begin{equation}
\frac{d}{dx} x \frac{d}{dx}\ln\tau \ \equiv \ 0 \ ,
\label{part2}
\end{equation}
also holds true. Using eq.~(\ref{part1}), we write
\begin{equation}
\ln\tau \ = \ \ln\frac{\beta_0^2}{2} \ - \ \ln x \ + \ 
\ln\frac{d}{dx}\left(x^2 f^2\right)\, ,
\label{part3}
\end{equation}
and substituting this form into eq.~(\ref{part2}) we find an ODE
for all the differential-rotation laws $f(x)$ that satisfy eq.~(\ref{part2}) 
identically and make eq.~(\ref{part1}) a family of exact solutions of
the Lane-Emden equation with rotation:
\begin{equation}
\frac{d}{dx} x \frac{d}{dx} \ln \frac{d}{dx}\left(x^2 f^2\right) \ = \ 0 \, .
\label{part4}
\end{equation}
This third-order linear ODE can be readily integrated to yield 
the following results:
\begin{equation}
\frac{d}{dx}\left(x^2 f^2\right) \ = \ A x^k \ ,
\label{class1}
\end{equation}
implying that
\begin{equation}
\tau (x) \ = \ \frac{\beta_0^2}{2}\cdot A x^{k-1} \ ,
\label{class2}
\end{equation}
and that
\begin{equation}
f(x) \ = \ \frac{\sqrt{A\cdot g(x) + B}}{x} \ ,
\label{class3}
\end{equation}
where $A$, $B$, and $k$ are arbitrary integration constants and
\begin{equation}
g(x) \ \equiv \  \left\{ \begin{array}{cc} 
         x^{k+1}/(k+1) \, , & \ {\rm if} \ \ \ k \neq -1 \\
         \ln x \, , \ \ \ \ \ \ \ \ \ \ \ & \ {\rm if} \ \ \ k = -1 
         \end{array} \right. \ ,
\label{class4}
\end{equation}
implying that ~$dg/dx = x^k$ ~for all values of $k$.
With so many free parameters ($A$, $B$, and $k$) in the differential-rotation 
profile, these solutions can easily become a theorist's playground. Here we highlight 
just a few interesting cases:

\begin{enumerate}
\item[(a)]{\it Parameter Constraints}.---Eq.~(\ref{class2}) shows that 
$\tau (x) > 0$ only for $A > 0$. This constraint also limits the physical
values of $k$ when $B \leq 0$ in eq.~(\ref{class3}); for example, 
$k\geq -1$ when $B=0$. This limitation can be easily circumvented by 
implementing composite rotation profiles with $B>0$ (see item (d)  and \S~\ref{comp} below).

\item[(b)]{\it Monotonically Decreasing Profiles}.---Eq.~(\ref{class2}) shows that
$\tau (x)$ is a decreasing function of $x$ for $k < 1$. The same condition
is sufficient to also make $f(x)$ a decreasing function of $x$ provided that
$B\geq 0$ in eq.~(\ref{class3}).

\item[(c)]{\it Uniform Rotation}.---For $A=2$, $B=0$, and $k=1$, eq.~(\ref{class3})
reduces to $f(x)=1$ and the equilibrium density (eq.~(\ref{class2})) then is
~$\tau (x) = \beta_0^2 = {\rm constant}$. Note that this constant cannot be
adjusted freely, and this is the reason for the oscillatory density profile in a uniformly rotating polytropic model found by \cite{jea14} and later by \cite{rob68}. (This solution is effectively a zeroth-order Bessel function and it oscillates in a regular manner about the $\tau=\beta_0^2$ intrinsic solution provided by the uniform rotation.)

\item[(d)]{\it Composite Profiles with} $B>0$.---Steep density profiles 
with $k < -1$ can be obtained by selecting $B > 0$ and by incorporating a 
central core region in uniform rotation. 
Even more complex equilibrium profiles
can be constructed by combining two or more density power laws (see \S~\ref{comp}).

\item[(e)]{\it Asymptotic Regime}.---For $k < -1$ and $B>0$, eq.~(\ref{class4})
shows that $g(x)\to 0$ as $x\to\infty$ and eq.~(\ref{class3}) then exhibits  the
asymptotic behavior $f(x)\to \sqrt{B}/x$. Therefore, all steep density profiles
with $k < -1$ and $\tau (x)\propto x^{k-1}$ approach a flat rotation curve 
($\Omega R\to$ constant) from above at large radii, independently of the value of $k$. Such a flat rotation curve has been recently observed in the Class 0 young protostellar system HH 211-mms in Perseus \citep{lee18}. 
\end{enumerate}

\subsection{Physical Interpretation}

From the perspective of the physics that dictates the above profiles, 
the solutions~(\ref{part1}) of the Lane-Emden equation~(\ref{main1}) describe a 
class of differentially rotating self-gravitating equilibria in which $z$-gradients are neglected
and the {\it radial gradient} of the 
gravitational acceleration is balanced exactly by the {\it radial gradient} 
of the centrifugal acceleration at every radius $x$. This occurs because, in the isothermal Lane-Emden
equation, we have gone to second order by taking an extra derivative on the components of the 
equation of hydrostatic equilibrium. The balance of gradients can be seen, most easily, by 
substituting eq.~(\ref{part1}) into the one-dimensional Poisson's equation ${\bf\nabla}^2\psi = \tau$, 
where $\psi\equiv\Phi / c_0^2$ is the normalized potential; the result is
\begin{equation}
\frac{1}{x}\frac{d}{dx} x \left[\frac{d\psi}{dx}\right]
\ = \ \
\frac{1}{x}\frac{d}{dx} x \left[\frac{1}{2}\beta_0^2 \cdot x f^2\right] 
\, .
\label{gradients}
\end{equation}
In this equation, the bracketed terms are the gravitational and centrifugal accelerations, 
respectively. This type of balance is different than the hydrostatic balance commonly discussed between the
magnitudes of these two accelerations in rotating gravitating systems;
and the power-law density solutions are borne out of this conformance of 
the two gradients. In the isothermal gaseous case of interest here, a
power-law density profile satisfies naturally the condition that the {\it radial variation}
of the enthalpy gradient ~$\rho^{-1}(dP /d\ln R)$ ~be zero (see eq.~(\ref{part2})) 
and so the pure power-law profile is not at all influenced by the radial variation of the 
pressure gradient---it becomes an exact intrinsic solution of eq.~(\ref{main1}).

\begin{figure}
\begin{center}
    \leavevmode
      \includegraphics[trim=0 0cm 0 0.1cm, clip, angle=0,width=9 cm]{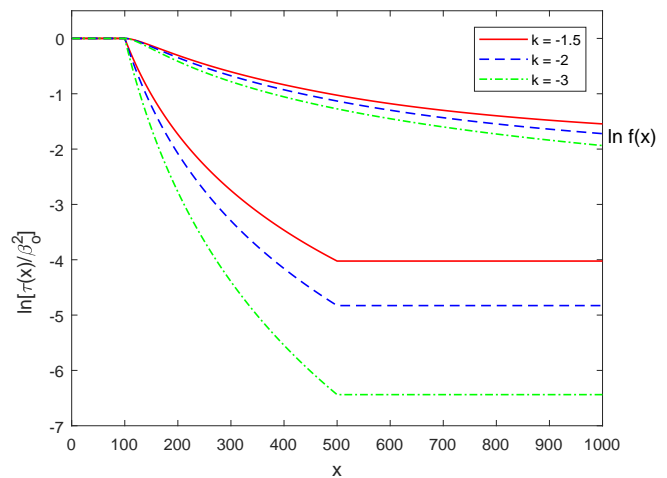}
\caption{Analytic density and rotation profiles of composite equilibrium models
for $x_1=100$, $x_2=500$, and $k = -1.5, -2$, and $-3$. The density
profile $\tau(x)$ is uniform for $x\leq x_1$ and for $x\geq x_2$; and
it follows the power law ~$x^{k-1}$ ~in the in-between region.
The rotation profile $f(x)$ is uniform for $x\leq x_1$ and monotonically
decreasing for $x > x_1$.
\label{fig1}}
  \end{center}
\end{figure}

\begin{figure}
\begin{center}
    \leavevmode
      \includegraphics[trim=0.5cm 1.75cm 0.5cm 1.75cm, clip, angle=0,width=9 cm]{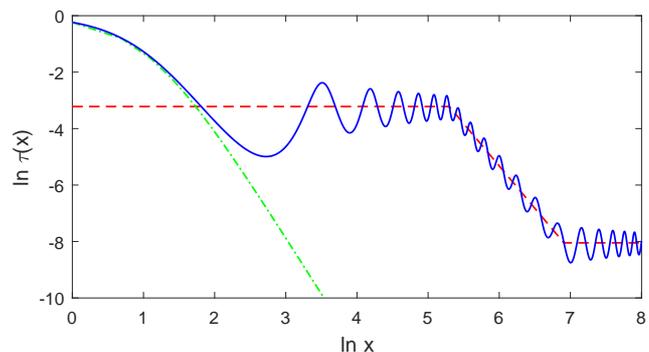}
\caption{Equilibrium density profile for a model with rotation parameter $\beta_0 = 0.2$ and a composite rotation profile with $x_1=200$, $x_2=1000$, and $k = -2$. 
The Cauchy solution (solid line) is forced to oscillate permanently about the intrinsic solution (dashed line) of the Lane--Emden ODE. The nonrotating 
analytical solution (dash-dotted line) is also shown for reference.
\label{fig2}}
  \end{center}
\end{figure}

\subsection{Composite Equilibrium Profiles}\label{comp}

Uniform rotation produces nearly equidistant peaks in the density profile. Such equidistant locations are also found in the inner three planets of our solar system and in the outer planets beyond Jupiter, and they are responsible for invalidating the Titius-Bode rule \citep{chr07,chr17}. A uniformly rotating inner core is also necessary to avoid the singularity of the power-law density profile at $x=0$. We construct then composite equilibria that incorporate an inner and an outer flat region in the density profile. Fig.~\ref{fig1} shows three such intrinsic solutions\footnote{Intrinsic singular solutions that do not depend on boundary conditions have been found in many other ODEs and they have often been called ``trivial.'' As we shall see here, they are extremely important in that they fully determine the behavior of the solutions of the Cauchy problem when boundary conditions are externally imposed.} of eq.~(\ref{main1}) for slopes of $k = -1.5, -2$, and $-3$.

In Fig.~\ref{fig2}, we show a numerical solution of the Cauchy problem with the usual boundary conditions for a centrally condensed model: $\tau(0)=1$ and $[d\tau/dx](0)=1$. The integrations were performed with \textsc{Matlab} \citep{sha97,sha99}.
The rotating intrinsic solution and the nonrotating solution are also plotted for comparison purposes. The two rotating solutions share the same differential rotation profile \citep[for more details, see][]{chr07}. The following features stand out in the figure: 
\begin{itemize}
\item[(a)] At small radii, the numerical (Cauchy) solution starts out very close to the nonrotating solution. 
\item[(b)] As soon as it crosses below the intrinsic solution, it gets attracted, turns around, and tries to match it. But the Cauchy solution cannot match the intrinsic solution because of the imposed boundary conditions. Thus, it is forced to oscillate permanently about the intrinsic solution, and this is how the many density maxima are created. 
\item[(c)] Density maxima are approximately equidistant in the inner and outer regions, where the intrinsic density profile is flat, although the equilibrium disk is still differentially rotating. 
\end{itemize}

\subsection{A Model of the Early Isothermal Solar Nebula}

\begin{figure}
\begin{center}
    \leavevmode
      \includegraphics[trim=0.5cm 1.75cm 0.5cm 1.75cm, clip, angle=0,width=9 cm]{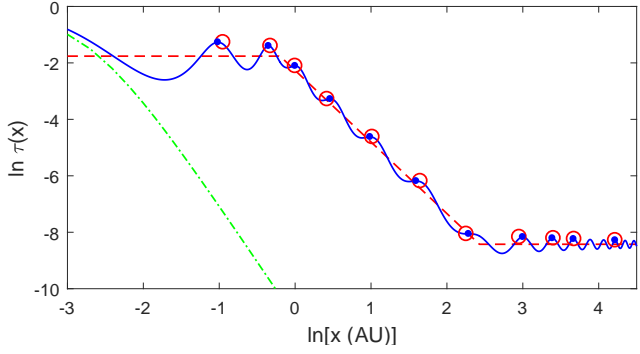}
\caption{Equilibrium density profile for the midplane of the solar 
nebula. Key: E(Earth), C(Ceres), J(Jupiter), P(Pluto), E(Eris). The Cauchy solution (solid line) has been fitted to the present solar system so that its density maxima (dots) correspond to the observed semimajor axes of the planetary orbits (open circles). The third density maximum was scaled to a distance of 1~AU. The mean relative error of the fit is 4.1\%, affirming that this simple equilibrium model produces an incomparable match to the 11 observed data points. The intrinsic solution (dashed line) and the nonrotating analytical solution (dash-dotted line) are also shown for reference. The last two planets shown in the fit are Pluto and Eris in the Kuiper belt.
\label{fig3}}
  \end{center}
\end{figure}

In Fig.~\ref{fig3}, we show the same model optimized to fit the locations of the planets in our solar system. Planets are expected to be found at density maxima, where the gravitational potential has minima in which dust and pebbles are expected to be trapped during the entire evolution of the solar nebula. The free parameters of the model and their best-fit values are: $k=-1.5$, $\beta_0=0.41$ (or, equivalently, the inner core radius $x_1=0.82$ AU), and the outer radius $x_2=11$ AU. These are determined by setting the location of the third density maximum (Earth) to 1 AU, which also fixes the radial scale length of the model to $R_0=0.022$ AU ($\simeq 4.7$ solar radii). This best-fit model of the solar nebula is slowly rotating and thus stable to nonaxisymmetric perturbations. Its rotation parameter $\beta_0=0.41$ is below the critical value\footnote{This critical value is obtained from the $\alpha$-parameter criterion for stability of rotating, self-gravitating, gaseous disks $\alpha\leq\alpha_*= 0.35$ \citep{chr95} by combining $\beta_0\equiv \Omega_0/\Omega_J$ with  $\alpha\simeq\Omega_0/(\Omega_J\sqrt{2})$ to get $\beta_*\simeq\alpha_*\sqrt{2}\simeq 0.50$.} of $\beta_*\simeq 0.50$ for the onset of nonaxisymmetric  instabilities (see also \S~\ref{slow} below). 

\subsubsection{Equation of State and Threshold Density for Planet Formation in the Solar Nebula}\label{rhomax}

Using the scale length of the disk ($R_0=0.022$~AU) 
in eq.~(\ref{length}), we can write the equation of state for the gas as
\begin{equation}
\frac{c_0^2}{\rho_0} \ = \ 4\pi G R_0^2 \ = \ 9\times 10^{16} 
{\rm ~cm}^5 {\rm ~g}^{-1} {\rm ~s}^{-2}\, ,
\label{crho}
\end{equation}
where $c_0$ and $\rho_0$ are the local sound speed and the local density in the inner disk, respectively.
For an isothermal gas at temperature $T$, ~$c_0^2 = {\cal R} T/\overline{\mu}$, 
~where $\overline{\mu}$ is the mean molecular weight and ${\cal R}$ is the
universal gas constant. Hence, eq.~(\ref{crho}) can be rewritten as
\begin{equation}
\rho_0 \ = \ 9\times 10^{-10}\left(\frac{T}{\overline{\mu}}\right) \
{\rm ~g} {\rm ~cm}^{-3}\, ,
\label{trho}
\end{equation}
where $T$ and $\overline{\mu}$ are measured in degrees Kelvin and 
${\rm ~g} {\rm ~mol}^{-1}$, respectively. 

For the coldest gas with $T \geq 10$~K 
and $\overline{\mu} = 2.34 {\rm ~g} {\rm ~mol}^{-1}$ (molecular hydrogen and
neutral helium with fractional abundances $X=0.70$ and $Y=0.28$ by
mass, respectively), we find that
\begin{equation}
\rho_0 \ \geq \ 4\times 10^{-9} \ {\rm ~g} {\rm ~cm}^{-3}\, .
\label{therho}
\end{equation}
This value is comfortably larger than the well-known threshold for planet formation in the solar nebula \citep[$\rho_*\simeq 10^{-9} {\rm ~g} {\rm ~cm}^{-3}$; see, e.g.,][]{lis93} and it implies that the conditions for planet formation were already in place, at least in the inner disk, during the early isothermal phase \citep{toh02} of the solar nebula.

\subsubsection{Rotational State of the Solar Nebula}\label{slow}

Using the characteristic density $\rho_0$ of the inner disk
(eq.~(\ref{therho})) in the definition of ~$\Omega_J\equiv\sqrt{2\pi G\rho_0}$, ~we
can determine the Jeans frequency of the disk:
\begin{equation}
\Omega_J \ = \ 4\times 10^{-8} {\rm ~rad} {\rm ~s}^{-1}\, .
\label{thej}
\end{equation}
Then, using the model's value $\beta_0 = 0.41$ in the definition 
of ~$\beta_0\equiv \Omega_0 /\Omega_J$, we can determine the angular velocity
of the uniformly-rotating core ($x_1\leq 0.82$~AU), viz.
\begin{equation}
\Omega_0 \ = \ 1.64\times 10^{-8} {\rm ~rad} {\rm ~s}^{-1}\, .
\label{theom}
\end{equation}
For reference, this value of $\Omega_0$ corresponds to an orbital period 
of 12~yr. In the present solar system, that would correspond to
a Keplerian orbit with semimajor axis ~$a = 5.2$~AU. Thus, the core of the solar nebula was rotating about as slowly as Jupiter is presently revolving around the Sun. This slow rotation of the core provides another indication that the solar nebula was not prone to nonaxisymmetric, self-gravity induced instabilities.

\section{Application to the Dark Gaps Discovered by ALMA}\label{apps}

\subsection{Preliminaries}

The above numerical solutions of the Cauchy problem have enjoyed more than ten years of obscurity as viable models of the solar nebula because of two main reasons, both linked to planet migration: (a) Virtually all of the exoplanetary systems discovered so far show gaseous giants that have migrated toward their central stars, having destroyed in the process any pre-existing, well-organized, inner planets. (b) In the solar system, an event called the Late Heavy Bombardment (LHB) has been hypothesized to have occurred $\sim$3.8 Gyr ago in an effort to explain cratering on the moon's surface. Migration of the gaseous giants and resonant interactions could explain such an event.

Both of these objections appear to be invalid nowadays: 
\begin{itemize}
\item[(a)] Recent ALMA, SPHERE, and Keck/NIRC2 observations show planets forming in circular orbits in young protoplanetary disks. We point out that no images show dark crossovers between the annular gaps, which would be an indication that migration is taking place \citep{alm15,and16,rua17,mac18,ave18,cla18,kep18,guz18,ise18,zha18,dul18,fav18,hua18,per18,kud18,lon18,pin18,vdm19}. \\ 
\item[(b)] The LHB has been recently disputed both on observational grounds and by simulations \citep{rey15,kai15,zel17,nes17,nes18,cle18}. The unfortunate fact that this hypothetical event was successfully modeled by many past simulations invalidates all of these previous modeling efforts altogether. Researchers in the field should really take notice of this fact, as unpleasant as its realization may be.
\end{itemize}
So it seems that the tide has turned and the answer to question (a) in \S~\ref{intro1} appears to be that the protoplanetary cores in our solar system formed in situ, safely inside gravitational potential minima in which dust and pebbles were trapped. None of these minima were in mean-motion resonance (MMR) (because of the particular radial density/rotation profiles) after the formation of the solar system concluded, and there were no significant perturbations to stir up the system. Thus, extensive planet migration does not appear to have occurred in our solar system, except for the limited radial excursions that the cores may have undertaken inside their local gravitational potential minima (of order of 1-3 AU for the outer gaseous giants).

Finally, we note that our model's intrinsic density profile argues against the Nice model \citep{tsi05,mor05,gom05} for the another reason as well: the giant protoplanets needed extended gravitational potential wells (of a few AU) in order to accumulate their large solid cores. Had they all formed in a compact configuration in an outer low-density region of the disk, there would have been insufficient amounts of planetesimals available for the process. In the model of Fig.~\ref{fig3}, the density drops by one order of magnitude in the region currently occupied by the gaseous giants, and beyond 10 AU, a slender annulus with a width of a few AU contains barely enough dust/planetesimals to form one giant core.

Based on the recent results produced by ALMA and SPHERE, we conclude that protoplanets in configurations that resemble the planet distribution in our solar system may not be found orbiting around nearby stars (a distance/volume limited sample anyway); instead our chances are much better with circumstellar disks around very young protostars. In such cases, we should check whether any dark annular gaps (a signature of planet formation) are located in MMRs; and whether the Titius-Bode rule plays any role in such young disks. Planets forming near MMRs are a source of future instability and the Titius-Bode rule \citep{chr17} is important in that observers use it in order to predict the probable locations of any yet undetected planets in multi-planet exosystems.

We note however that our investigation of MMRs continues to be valid in systems in which additional undetected planets have formed, although our study below will be incomplete. On the other hand, the analysis of the Titius-Bode rule becomes invalid in the presence of additional undetected planets. With these issues in mind, we proceed to analyze five well-organized ALMA disks with multiple circular, dark, well-organized gaps. 

\subsection{No Mean-Motion Resonances in the ALMA Disks}\label{mmrs}

1. {\it HL Tau}.---We investigated the dark rings of HL Tau \citep{alm15} for posible MMRs. 
We calculated all ratios $(a_i/a_j)^{3/2}$ ($i\neq j$) between the semimajor axes of the gaps and no pair of orbits appear to be in MMR. One ratio deviates from exact low-order commensurability by 7.4\% and the remaining ratios deviate by more than 10\%. Thus, this young solar system appears to be a survivor for ages to come. \\

\noindent
2. {\it CI Tau}.---\cite{cla18} have reported that the dark gaps of this system are not in a resonant configuration. Based on the relatively high disk mass (0.92~$M_\odot$), they speculated that migration may still take place in this environment. Of course, this is not going to occur, if the planets are residing in gravitational potential minima of the gas distribution. \\
 
\noindent
3. {\it AS 209}.---The latest ALMA observations \citep{guz18,hua18,zha18} show seven dark annular gaps in this system. We calculated all ratios $(a_i/a_j)^{3/2}$ ($i\neq j$) between the semimajor axes of the rings and no pair of orbits appear to be in MMR. One ratio deviates from exact low-order commensurability by 5.2\% and the remaining ratios deviate by more than 12\%. Thus, this young solar system also appears to be in a nonresonant configuration. \\

\noindent
4. {\it HD 163296}.---\cite{ise18} show four dark rings in this system. Elliptical fitting of the isophotes has produced precise measurements of their semimajor axes (9.96, 44.77, 86.61, and 140.62 AU). No pair of orbits appear to be in MMR. Two ratios deviate from exact low-order commensurability by 5.0\% and 6.9\% and the remaining ratios deviate by more than 31\%. This young solar system also appears to be in a nonresonant configuration. \\

\noindent
5. {\it TW Hya}.---ALMA and Keck/NIRC2 observations \citep{and16,rua17} indicate the presence of dark gaps at 1, 24, 41, 47, and 88 AU. Once again, no pair of orbits appear to be in MMR. A possible exception is the period ratio between the second and fifth dark gaps that may be thought to be in a 7:1 MMR (a deviation of 2.1\%, as opposed to all other deviations that are larger than 14\%). In this case as well, we cannot conclude that the system is in a resonant configuration. \\

We conclude that the above protoplanetary systems appear to be stable and they are not in danger of suffering resonant interactions/instabilities after the gas disks are gone. In this sense, these systems appear to be similar to our own solar system, not only because of their well-organized structures, but also because of their long-term stability properties. Thus, we argue that a population of protoplanetary systems roughly similar to the solar nebula has now been found by the latest high-resolution observations of ALMA/DSHARP disks; and this identification provides an answer to question (b) in \S~\ref{intro1} above.

\subsection{No Titius-Bode Arrangements in the ALMA Disks}

We also examined the dark rings of the above systems to find out if any successive pairs obey approximately the Titius-Bode rule. In such a case, that would imply that the differential rotation and radial density profiles would be similar to those of the solar nebula. The results show that none of these disks has similar physical characteristics to the solar nebula. 

We calculated a geometric progression of the form $a_i=\sqrt{a_{i-1}\cdot a_{i+1}}$ and also an arithmetic progression of the form $a_i=(a_{i-1} + a_{i+1})/2$ for a uniformly rotating profile, where $a_i$ is the semimajor axis of the $i^{\rm th}$ orbit. The smallest geometric deviations between  neighboring gaps range from 6\% to 39\% in the five systems listed in \S~\ref{mmrs}. Furthermore, a uniform rotation profile in the inner regions is also ruled out for the two systems each of which shows seven dark gaps (the smallest deviations are 8.3\% for AS 209 and 15\% for HL Tau, respectively).

But of course there could be more planets forming in these systems that are yet undetected. On the other hand, there is no known universal mechanism (neither one presently hypothesized) that would impose more or less the differential rotation profile of the solar nebula to these disks. Therefore, the above results arguing against the Titius-Bode rule also testify to the expected diversity of radial rotation/density/temperature profiles in young planet-forming protoplanetary nebulae such as the observed ALMA/DSHARP disks (\S~\ref{intro}).

\section{Summary}\label{disc}

The latest high-resolution observations from ALMA/DSHARP and SPHERE have produced images of protoplanetary disks in which 3-7 planets have already formed (\S~\ref{mmrs}) in well-organized and non-interacting dark gaps, a signature of planet formation very much unlike the picture which has been produced from exoplanet searches (\S~\ref{intro1}). There is no evidence of violent evolution in these systems, no dramatic planet migrations, no dominant nonaxisymmetric features, no self-gravity induced instabilities, and no stirring from streaming/radial motions in the gas. Unlike in extrasolar systems, nebular evolution has not concluded yet, the gas disks are still present and, for the most part, they are not in Keplerian rotation \citep[see, e.g.,][]{lee18}, although some form of differential rotation is expected to be pervasive in these systems anyway. 

We are then facing an enormous discrepancy between the end-product of planetary formation and the very early stages of ALMA-observed protoplanetary nebulae that occur in relatively massive gaseous disks still in the process of forming their central stars. This discrepancy deserves an explanation. 

As far as we know, there currently exists only one physical model of well-organized, ordered planet formation in gaseous isothermal disks. This model was described in \S~\ref{theory}. It is based on a formal mathematical property of nonlinear differential equations, such as the isothermal Lane-Emden equation with rotation that is applicable to our context. Such equations admit ``intrinsic'' (often called trivial) solutions that are independent of boundary conditions. Trivial or not, these solutions are preferred by the equations themselves and they determine all other physically relevant solutions (the Cauchy problem). This is now understood as follows: When boundary conditions are externally imposed, the resulting Cauchy problem does not conform to the preferences of the differential equation and produces new (physical) solutions that are either repelled from or attracted to the intrinsic solutions \citep{chr16a,chr16b,kat18}. Cauchy solutions attracted to an underlying intrinsic solution have no choice but to oscillate permanently about it, since the chosen boundary conditions forbid an exact match. 

The Lane-Emden equation with rotation that is relevant to the solar nebula is one of these equations whose Cauchy solutions are attracted to the intrinsic solution and they are forced to oscillate permanently about it. The corresponding equilibrium density profile develops density maxima (i.e., gravitational potential minima) which can trap dust and planetesimals very early on in the isothermal phase of evolution of real protostellar disks. Protoplanets then form in relative safety inside these potential minima and long before the central stars are formed \citep[see also][]{gre10}.

Virtually all protoplanetary disks observed by ALMA and SPHERE and during the DSHAR Project exhibit these characteristics, well-organized annular dark gaps with planets orbiting in circular orbits and in the complete absence of gravitational interactions, radial migrations, or self-gravitating instabilities (\S~\ref{apps}). This robust and novel mathematical model argues strongly against violent planet formation scenarios such as those speculated in the past (planet destruction via migration of gas giants, the Nice model, and instabilities caused by planet crossing of MMRs and/or disk self-gravity). We believe that future observations will only add support to this simple but fundamental physical picture of early planet formation in protoplanetary disks during their early phase of isothermal evolution. We are in the process of modeling the ALMA disks to learn about the physical conditions of these planet-forming young systems.

\end{document}